\documentclass[longauth]{aa}
\usepackage{url} 
\usepackage{epstopdf}
\usepackage{color}
\usepackage{gensymb}
\usepackage{textcomp}
\usepackage{multirow}
\usepackage{ragged2e}
\usepackage{hyperref}
\usepackage{amsmath} 
\usepackage{booktabs}
\usepackage{caption}
\usepackage{afterpage}
\usepackage{mathtools}
\usepackage{tabularx}
\usepackage{bold-extra}
\usepackage{xspace}
\usepackage{relsize}
\usepackage[utf8]{inputenc}
\usepackage{subcaption}
\usepackage{graphicx}
\usepackage{soul}
\usepackage[normalem]{ulem}
\usepackage{changes}

\usepackage[acronym,hyperfirst=false]{glossaries}
\glsenableentrycount 
\newacronym{SNR}{SNR}{supernova remnant}
\newacronym{LAT}{LAT}{Large Area Telescope}
\newacronym{MAGIC}{MAGIC}{Major Atmospheric Gamma Imaging Cherenkov}
\newacronym{IACT}{IACT}{Imaging Atmospheric Cherenkov Telescope}
\newacronym{MARS}{MARS}{MAGIC Analysis and Reconstruction Software}
\newacronym{FoV}{FoV}{field of view}
\glsdisablehyper

\def\fermi{\textit{Fermi}\xspace}
\def\fermilat{\textit{Fermi}-LAT\xspace}

\def\w44{W44\xspace} 
\def\W44{W44\xspace}

\newcommand{\h}{$^{\rm h}$}
\newcommand{\m}{$^{\rm m}$}

\newcommand{\dotdeg}{\rlap{.}^\circ}

\begin{document}

\title{Cosmic-ray acceleration and escape from supernova remnant \w44 as probed by \fermilat and MAGIC}
\titlerunning{W 44 with \fermilat and MAGIC}

%
\author{
\footnotesize
%
S.~Abe\inst{1} \and
J.~Abhir\inst{2} \and
A.~Abhishek\inst{3} \and
V.~A.~Acciari\inst{4} \and
A.~Aguasca-Cabot\inst{5} \and
I.~Agudo\inst{6} \and
T.~Aniello\inst{7} \and
S.~Ansoldi\inst{8,44} \and
L.~A.~Antonelli\inst{7} \and
A.~Arbet Engels\inst{9}
% ...
}
\institute {Japanese MAGIC Group: Institute for Cosmic Ray Research (ICRR), The University of Tokyo, Kashiwa, 277-8582 Chiba, Japan
\and ETH Z\"urich, CH-8093 Z\"urich, Switzerland
\and Universit\`a di Siena and INFN Pisa, I-53100 Siena, Italy
\and Instituto de Astrof\'isica de Canarias and Dpto. de  Astrof\'isica, Universidad de La Laguna, E-38200, La Laguna, Tenerife, Spain
\and Universitat de Barcelona, ICCUB, IEEC-UB, E-08028 Barcelona, Spain
\and Instituto de Astrof\'isica de Andaluc\'ia-CSIC, Glorieta de la Astronom\'ia s/n, 18008, Granada, Spain
\and National Institute for Astrophysics (INAF), I-00136 Rome, Italy
\and Universit\`a di Udine and INFN Trieste, I-33100 Udine, Italy
\and Max-Planck-Institut f\"ur Physik, D-85748 Garching, Germany
\and Universit\`a di Padova and INFN, I-35131 Padova, Italy
\and Croatian MAGIC Group: University of Zagreb, Faculty of Electrical Engineering and Computing (FER), 10000 Zagreb, Croatia
\and IPARCOS Institute and EMFTEL Department, Universidad Complutense de Madrid, E-28040 Madrid, Spain
\and Centro Brasileiro de Pesquisas F\'isicas (CBPF), 22290-180 URCA, Rio de Janeiro (RJ), Brazil
\and University of Lodz, Faculty of Physics and Applied Informatics, Department of Astrophysics, 90-236 Lodz, Poland
\and Centro de Investigaciones Energ\'eticas, Medioambientales y Tecnol\'ogicas, E-28040 Madrid, Spain
\and Institut de F\'isica d'Altes Energies (IFAE), The Barcelona Institute of Science and Technology (BIST), E-08193 Bellaterra (Barcelona), Spain
\and Departament de F\'isica, and CERES-IEEC, Universitat Aut\`onoma de Barcelona, E-08193 Bellaterra, Spain
\and Universit\`a di Pisa and INFN Pisa, I-56126 Pisa, Italy
\and INFN MAGIC Group: INFN Sezione di Bari and Dipartimento Interateneo di Fisica dell'Universit\`a e del Politecnico di Bari, I-70125 Bari, Italy
\and Armenian MAGIC Group: A. Alikhanyan National Science Laboratory, 0036 Yerevan, Armenia
\and Department for Physics and Technology, University of Bergen, Norway
\and INFN MAGIC Group: INFN Sezione di Torino and Universit\`a degli Studi di Torino, I-10125 Torino, Italy
\and INFN MAGIC Group: INFN Sezione di Catania and Dipartimento di Fisica e Astronomia, University of Catania, I-95123 Catania, Italy
\and Croatian MAGIC Group: University of Rijeka, Faculty of Physics, 51000 Rijeka, Croatia
\and Universit\"at W\"urzburg, D-97074 W\"urzburg, Germany
\and Technische Universit\"at Dortmund, D-44221 Dortmund, Germany
\\
\\
\and University of Geneva, Chemin d'Ecogia 16, CH-1290 Versoix, Switzerland 
\and Japanese MAGIC Group: Physics Program, Graduate School of Advanced Science and Engineering, Hiroshima University, 739-8526 Hiroshima, Japan
\and Deutsches Elektronen-Synchrotron (DESY), D-15738 Zeuthen, Germany
\and Armenian MAGIC Group: ICRANet-Armenia, 0019 Yerevan, Armenia
\and Croatian MAGIC Group: University of Split, Faculty of Electrical Engineering, Mechanical Engineering and Naval Architecture (FESB), 21000 Split, Croatia
\and Croatian MAGIC Group: Josip Juraj Strossmayer University of Osijek, Department of Physics, 31000 Osijek, Croatia
\and Finnish MAGIC Group: Finnish Centre for Astronomy with ESO, Department of Physics and Astronomy, University of Turku, FI-20014 Turku, Finland
\and Japanese MAGIC Group: Department of Physics, Tokai University, Hiratsuka, 259-1292 Kanagawa, Japan
\and Saha Institute of Nuclear Physics, A CI of Homi Bhabha National Institute, Kolkata 700064, West Bengal, India
\and Inst. for Nucl. Research and Nucl. Energy, Bulgarian Academy of Sciences, BG-1784 Sofia, Bulgaria
\and Japanese MAGIC Group: Department of Physics, Yamagata University, Yamagata 990-8560, Japan
\and Finnish MAGIC Group: Space Physics and Astronomy Research Unit, University of Oulu, FI-90014 Oulu, Finland
\and Japanese MAGIC Group: Chiba University, ICEHAP, 263-8522 Chiba, Japan
\and Japanese MAGIC Group: Institute for Space-Earth Environmental Research and Kobayashi-Maskawa Institute for the Origin of Particles and the Universe, Nagoya University, 464-6801 Nagoya, Japan
\and Japanese MAGIC Group: Department of Physics, Kyoto University, 606-8502 Kyoto, Japan
\and INFN MAGIC Group: INFN Roma Tor Vergata, I-00133 Roma, Italy
\and Japanese MAGIC Group: Department of Physics, Konan University, Kobe, Hyogo 658-8501, Japan
\and also at International Center for Relativistic Astrophysics (ICRA), Rome, Italy
% ...
}

\authorrunning{MAGIC Collaboration}

\date{\today}

\abstract  {
The supernova remnant (SNR) \W44 and its surroundings are a prime target for studying the acceleration of cosmic rays (CRs). Several previous studies established an extended gamma-ray emission that is set apart from the radio shell of \W44. This emission is thought to originate from escaped high-energy CRs that interact with a surrounding dense molecular cloud complex.
}
{
We present a detailed analysis of \fermilat data with an emphasis on the spatial and spectral properties of \W44 and its surroundings. We also report the results of the observations performed with the MAGIC telescopes of the northwestern region of \W44. Finally, we present an interpretation model to explain the gamma-ray emission of the SNR and its surroundings.
}
{
We first performed a detailed spatial analysis of 12\,years of \fermilat data at energies above 1\,GeV, in order to exploit the better angular resolution, while we set a threshold of 100\,MeV for the spectral analysis. We performed a likelihood analysis of 174\,hours of MAGIC data above 130\,GeV using the spatial information obtained with \fermilat . 
}
{
The combined spectra of \fermilat and MAGIC, extending from 100\,MeV to several TeV, were used to derive constraints on the escape of CRs.
Using a time-dependent model to describe the particle acceleration and escape from the SNR, we show that the maximum energy of the accelerated particles has to be $\simeq 40$\,GeV. However, our gamma-ray data suggest that a small number of lower-energy particles also needs to escape. We propose a novel model, the broken-shock scenario, to account for this effect and explain the gamma-ray emission.}
{}
\maketitle

\section{Introduction}
\label{sec:intro}

One of the main topics in astroparticle physics is the identification of the sources of cosmic rays (CRs). CRs are particles that originate in astrophysical environments and have energies up to $\sim 10^{20}$eV. Since the second half of the last century \citep{ginz_syr_1964}, supernova remnants (SNRs) have been considered the sources of the bulk of Galactic CRs. This hypothesis has indirectly been supported by recent observations of SNRs at MeV, GeV, and TeV gamma rays. In particular, a striking piece of evidence is the so-called pion-decay bump, which can be observed in the energy range 50 MeV--200 MeV. This spectral feature unequivocally identifies gamma rays produced by $\pi^{0}$ decay, and consequently, relativistic protons \citep{Ackermann_2013}.

The SNR paradigm for the origin of CRs also received support from other observational evidence, including the detection of nonthermal X-ray filaments \citep{Vink:2012} as well as the broadening of H$\alpha$ lines from Balmer shocks \citep{Morlino:2014}. However, several aspects of the problem still remain unsolved. The most relevant aspect concerns the maximum energy: It is unclear from a theoretical perspective whether SNRs can explain the CR spectrum up to the knee energy of $\sim$\,PeV, as required by observations \cite[e.g.][]{Schure-Bell:2013, Cardillo+2015, Cristofari+2020}. From an observational point of view, the Large High Altitude Air Shower Observatory (LHAASO) collaboration has shown that only a few SNRs are associated with gamma-ray emission beyond $\sim 100$\,TeV \citep{Cao_2024}. Unexpectedly, they are all middle-aged SNRs, while younger objects seem not to show the same high-energy emission. Whether this emission results from particles that escaped from the SNRs when they were younger but are still located around the acceleration region, or if this is connected to pulsar wind nebulae \cite[e.g.][]{breuhaus2022} remains to be addressed. 
Moreover, the final CR spectrum that is released into the Galaxy is poorly constrained. It is in general different from the spectrum that is accelerated at the shock. Both aspects, the maximum energy and the released spectrum, are connected to the way in which particles escape from the SNR, which is difficult to describe given the high nonlinearity that is usually introduced in diffusive shock acceleration (DSA) theory \cite[see, e.g.][]{Blasi:2013}.  
One possible way to study the escape process is to examine middle-aged SNRs and try to detect gamma-ray emission from escaping particles.

In this regard, the case of the middle-aged SNR \W44 ($\sim$20\,kyr) is of great interest. It is embedded in a dense molecular cloud (MC), as observed, for example, by \citet{reach_shocked_2005} in the millimeter and near-infrared bands and by \citet{park_h_2013} in the HI 21 cm line. It was discovered at gamma-ray energies by \citet{w44_fermi_abdo_2010} and \citet{giuliani_neutral_2011} with the \textit{Fermi}-Large Area Telescope (LAT) and the Astrorivelatore Gamma ad Immagini LEggero (AGILE) Gamma-Ray Imaging Detector (GRID) instruments, respectively. \citet{Ackermann_2013} confirmed the finding that the low-energy cutoff in the gamma-ray spectrum is consistent with emission from decaying $\pi^0$ produced in collisions of accelerated protons. In addition, \citet{uchiyama_fermi-lat_2012}  showed the presence of gamma-ray emission toward the northwest (NW) and southeast (SE) regions of \W44, which has been interpreted as a sign of escaping CRs illuminating the MC complex.
More recently, \citet{peron_gamma-ray_2020} repeated the \fermilat analysis of the \w44 region with a threefold increased exposure compared to the previous study. They found the NW source to be significantly larger than previously thought. This might be caused by a large underlying background component, however, that is not properly included in the background model.

Since \w44 is a very bright source in the GeV band, its morphological features affect the features of the surrounding sources. Consequently, it is necessary to describe the morphology of the remnant well. 

Although the origin of gamma rays from \w44 has been established to be hadronic, the underlying acceleration mechanism is still debated. In its test-particle limit, the DSA model exhibits shortcomings as it tends to predict spectra that are harder than the observed spectra. Thus, following the original idea by \cite{Blandford-Cowie:1982}, numerous authors have proposed an alternative scenario in which preexisting Galactic CRs are reaccelerated and compressed by the expansion of the SNR forward shock into a dense environment \citep{Uchiyama+2010, Lee-Patnaude+2015, tang_time-dependent_2015, Cardillo+2016}. These models usually require a large fraction of the shock surface ($\gtrsim 50\%$) to interact with the dense molecular environment \citep{Cardillo+2016}. However, it was recently pointed out that some of these models require a circumstellar density that does not fit the evolutionary stage of the SNR \citep{deOnaWilhelmi+2020, Sushch-Brose:2023}.

Another plausible scenario is connected with the escape process. Spectra steeper than $E^{-2}$ can result from regular DSA when the time-dependent escape of particles is taken into account together with the fact that the maximum energy of accelerated particles is a decreasing function of time \citep{Celli:2019,Brose+2020,MAGIC_gamma-Cygni:2023}.  Models accounting for particle escape are also better suited to explain the gamma-ray emission from the region outside the SNRs. The escaping flux is tightly connected to the spectrum inside the SNRs, and when the latter is known, the former can therefore be also determined. In contrast, reacceleration models do not make clear predictions of the escaping flux.

The paper is organized as follows. In Sect. \ref{sec:obs_fermi} we present our analysis of \fermilat data and discuss the morphology (Sect. \ref{sec:morph_analysis}) and spectrum (Sect. \ref{sec:spect_analysis}). In Sect. \ref{sec:obs_magic} we present the novel \gls{MAGIC} observations and data analysis, and in Sect. \ref{sec:model}, we use the escape scenario to model the gamma-ray emission from \W44 and the surrounding regions. We summarize our results in Sect. \ref{sec:summary}.

\section{\fermilat observations and data analysis}
\label{sec:obs_fermi}

The \textit{Fermi Gamma-ray Space Telescope} is a gamma-ray observatory that was launched by NASA on June 11, 2008. It orbits at an altitude of $\sim$565 km with an inclination of 25$\dotdeg$6 with respect to the equator. The main instrument on board the satellite is the Large Area Telescope, which is an imaging pair-tracking telescope with a wide field of view that detects photons from 20 MeV to a few TeV \citep{LAT_description_2009}.

We selected data covering 142 months from a 15$^{\circ}$ region of interest (RoI) centered on the position of SNR \w44 (l: 34$\dotdeg$65, b: $-0\dotdeg38$) in the energy interval between 100 MeV and 2 TeV. 

We performed a binned likelihood analysis in a square region inscribed in the selected RoI. The data were spatially binned with pixels of 0$\dotdeg$1 and divided into energy bins using eight bins/decade. We selected the \texttt{SOURCE} event class, which is the recommended selection for steady point-like or moderately extended sources.
In order to increase the \fermilat sensitivity, we divided the dataset according to the event types available in the \texttt{Pass\ 8} reprocessing of \fermilat data \citep{Ajello_2021,Atwood_2013_pass8, bruel2018fermilat}. In particular, we used the point spread function (PSF) event types, which from PSF0 to PSF3 provide an increasingly better angular reconstruction.
We defined components by applying different cuts in event type, energy, and maximum zenith angle $z_{max}$. The latter was used to reduce the contamination from the Earth's limb. A tighter cut was applied at lower energy because the Earth's limb flux and the LAT PSF increase.

We combined the components using a summed likelihood analysis procedure in which we applied the corresponding set of IRFs, \texttt{P8R3\_SOURCE\_V3}, to each component.
We adopted the following components, as prescribed by \citet{abdollahi_fermi_2020}:
\begin{itemize}
	\item \texttt{PSF2} and \texttt{PSF3} events with $z_{max}$ = 90$^{\circ}$ for the energy range 100 MeV - 300 MeV;
	\item \texttt{PSF1}, \texttt{PSF2} and \texttt{PSF3} events with $z_{max}$ = 100$^{\circ}$ for the energy range 300 MeV - 1 GeV;
	\item \texttt{PSF0}, \texttt{PSF1}, \texttt{PSF2} and \texttt{PSF3} events with $z_{max}$ = 105$^{\circ}$ for the energy range 1 GeV - 2 TeV.
\end{itemize} 

We conducted a morphological analysis of the \w44 region by selecting \texttt{PSF2} and \texttt{PSF3} events with an energy above 1 GeV and with a maximum zenith angle of 105$^{\circ}$ in order to exploit the best angular resolution of the instrument\footnote{The 68\% containment angle of the acceptance-weighted PSF varies from $\sim$5$^{\circ}$ at 100 MeV to $\sim$0.5$^{\circ}$ at 1 GeV for \texttt{PSF2}+\texttt{PSF3} events}. On the other hand, we considered the full statistics in the entire energy range 100 MeV - 2 TeV to perform a spectral analysis. The correction for the energy dispersion was enabled.

We modeled the RoI considering all the known sources from the 4FGL-DR2 catalog \citep{4FGL_DR2_ballet2020fermi} within 20$^{\circ}$ from the RoI center and the Galactic and isotropic diffuse background models (\texttt{gll\_iem\_v07}, \texttt{iso\_P8R3\_SOURCE\_V3\_v1} \footnote{See here \url{https://fermi.gsfc.nasa.gov/ssc/data/access/lat/BackgroundModels.html} for more details.}). The normalization parameters of the diffuse components and of sources with a significance higher than 4 $\sigma$ were fit, while other parameters were kept fixed at the catalog values. Moreover, the correction for energy dispersion was disabled for the isotropic template\footnote{See here \url{https://fermi.gsfc.nasa.gov/ssc/data/analysis/documentation/Pass8_edisp_usage.html} for more details}.
We conducted the analysis using the fermitools v2.0.8 \citep{fermitools_2019} and the fermipy v1.0.1 \citep{Wood_2017_fermipy} packages. 

\subsection{Morphological analysis}
\label{sec:morph_analysis}
We first performed a detailed study of the  morphology and spectral shape of the region around the \W44 remnant in the energy range 1 GeV - 2 TeV. As already stated in the previous sections, we limited the energy range and selected the \texttt{PSF2}+\texttt{PSF3} event types in order to exploit the better PSF of the instrument. 

We first removed all known sources within 1$^{\circ}$ from the RoI center except for \w44. These were
\begin{itemize}
	\item 4FGL J1856.7+0125c, with flags\footnote{The analysis flags are used to indicate possible issues noted in the detection or characterization of the source. Flag n.3 indicates that the energy flux ($>$100 MeV) changed by more than 3$\sigma$ when changing the diffuse model or the analysis method, flag n.5 marks sources which are close to brighter neighbors, flag n.6 sources likely contaminated by the diffuse emission, flag n.9 identifies sources with bad localization 
    and flag n.12 sources with a highly curved spectrum. See \cite{abdollahi_fermi_2020} for full explanation.  } 3, 5, 6 ;
	\item 4FGL J1857.4+0106, with flags 5, 12;
	\item 4FGL J1855.8+0150, with flag 3;
	\item 4FGL J1857.1+0056, with flag 3;
	\item 4FGL J1857.6+0143, with flags 9,12;
	\item 4FGL J1854.7+0153, with flag 3;
	\item 4FGL J1859.3+0058, (no flags);
	\item 4FGL J1857.6+0212, with flag 5;
	\item 4FGL J1858.3+0209, with flag 5.
\end{itemize}

We then applied an algorithm designed to detect point-like sources within 1$^{\circ}$ from the center of the RoI. The algorithm used a test source with a \texttt{power-law} (PL) spectrum
    \begin{equation*}
       \text{(PL)} \quad \frac{dN}{dE}=N_{0}\left(\frac{E}{E_{0}}\right)^{-\gamma}
    \end{equation*}
with the source position (two parameters) and the normalization and index of the spectrum left free in the fit. The test statistics (TS) was calculated for the test source, where the TS is defined as $TS=2(\ln{\mathcal{L}_1}-\ln{\mathcal{L}_0})$, with $\ln{\mathcal{L}_1}$ and $\ln{\mathcal{L}_0}$ being the log-likelihood of the models with and without the test source, respectively.
Every source found with $TS>25$ (equivalent to a significance of approximately 4$\sigma$ for a fit with four degrees of freedom \citep{mattox_1996} was tested for evidence of spatial extension and spectral curvature, starting with the most significant source, and the best-fit model for that source was included in subsequent passes of the source-finding algorithm. The procedure was repeated until no new sources were found.
The spatial extension test was performed by refitting that source with a disk model in which the radial parameter was left free to vary.
A source was considered extended when $TS_{ext}>$ 25, where $TS_{ext}$ is defined as twice the log-likelihood difference between a model H1 with an extended disk model and a null hypothesis H0 based on a point-like source: $TS_{ext} = 2\left(\ln{\mathcal{L}_{disk}}-\ln{\mathcal{L}_{point}}\right)$. 
The sources were tested for spectral curvature by replacing the PL shape with a curved spectrum. As alternative models, a \texttt{log-parabola} (LP) and a \texttt{broken power-law} (BPL) spectrum \footnote{\url{https://fermi.gsfc.nasa.gov/ssc/data/analysis/scitools/source_models.html}}
    \begin{equation*}
        \begin{split}
            \text{(LP)} \quad \frac{dN}{dE}&=N_{0}\left(\frac{E}{E_{b}}\right)^{-\left(\alpha+\beta\ln\left( E/E_{b}\right)\right)}  \\ 
            \text{(BPL)} \quad \frac{dN}{dE}&=N_{0}\times 
                \begin{cases} 
                   \left(\frac{E}{E_{b}}\right)^{\gamma_{1}}     & \text{if } E < E_{b} 	\\
                   \left(\frac{E}{E_{b}}\right)^{\gamma_{2}}     & \text{otherwise }
                \end{cases}
        \end{split}
    \end{equation*}
were considered.
We evaluated the significance of the curvature using the quantity $TS_{curv} = 2\left(\ln{\mathcal{L}_{LP/BPL}}-\ln{\mathcal{L}_{PL}}\right)$. The quantity $\ln{\mathcal{L}_{LP/BPL}}$ represents the log-likelihood obtained with the LP/BPL model and $\ln{\mathcal{L}_{PL}}$ the one obtained with a PL model. The source was considered curved for $TS_{curv}>20$, and the proper spectral model was chosen as the one providing the highest value of $TS_{curv}$. 

We point out that the thresholds we used to define an extended or curved source are higher than those adopted in previous works (\cite{Ackermann_2018}, \cite{abdollahi_fermi_2020}). This was chosen to avoid an excessive increase in the number of free parameters in the model and to minimize the risk of source confusion.

In order to minimize the influence of close-by extended sources, we also performed an extension and spectral fit for Kes79 and HESS J1857+026, which lie beyond the one-degree circular region.

\subsection{\W44 morphology}
\label{sec:w44_morph_analysis}

The morphology of the extended SNR \w44 has been widely studied in gamma rays for almost 12 years. The first template, an elliptical ring, was derived by \citet{w44_fermi_abdo_2010} based on 11 months of data collected by the \fermilat. It is still used as the spatial model for this source in the  \fermi catalog \citep{4FGL_DR2_ballet2020fermi}. \citet{giuliani_neutral_2011} confirmed based on AGILE data the quasi-elliptical morphology of the emission, which resembles the 324 MHz radio continuum flux density detected by the Very Large Array (VLA) \citep{castelletti_2007}. Using three years of data of \fermilat, \citet{uchiyama_fermi-lat_2012} modeled the spatial distribution of the gamma-ray emission from \w44 with a 10\,GHz synchrotron radio map. \citet{Cardillo+2014} studied the \w44 SNR with AGILE data obtained up to June 2012, compared with VLA data and NANTEN2 telescope carbon monoxide (CO) data in the velocity channels 41\,km/s and 43\,km/s. They  highlighted that most of the gamma-ray emission comes from a site of interaction between the SNR and the MC. Finally, \citet{peron_gamma-ray_2020} showed in their analysis of almost 10 years of \fermilat data that the \w44 SNR morphology in gamma rays can be better described by an elliptical ring than with other spatial shapes they tested (disk, full ellipse, or radio templates derived from \cite{Condon_1998} and \cite{egron_2017}).

In order to determine the best spatial description of \w 44 SNR, we repeated the procedure illustrated in the previous section several times and each time adopted a different spatial template for \w 44. 
The \w 44 spectral shape was always described using an LP function, whose parameters were left free in the fitting procedure.  
We chose the best configuration by evaluating the value of the Akaike information criterion (AIC) \citep{Akaike_1998} for each analysis result. It is defined as $AIC=2k-2\ln{\mathcal{L}}$, where \textit{k} is the number of free parameters, and $\ln\mathcal{L}$ is the log-likelihood of the model. The best model is the one for which the AIC is minimized. Since the variation in the number of parameters depends essentially on what happens within 1$^{\circ}$ from the center of the RoI, we only considered the sources in this restricted region to calculate \textit{k}.

\begin{figure}
    \centering
    \begin{subfigure}[tb]{0.175\textwidth}
        {\includegraphics[width=\textwidth]{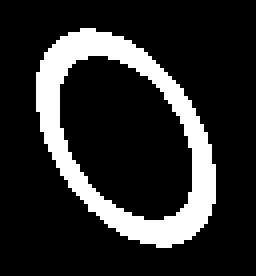}}
        \caption{}
    \end{subfigure}
    \begin{subfigure}[tb]{0.175\textwidth}
        \includegraphics[width=\textwidth]{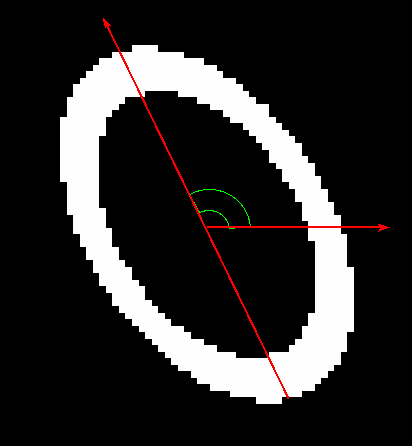}
        \caption{}
    \end{subfigure}\\
    \begin{subfigure}[tb]{0.175\textwidth}
        \includegraphics[width=\textwidth]{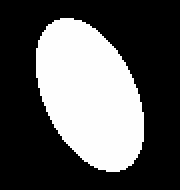}
        \caption{}
    \end{subfigure}
    \begin{subfigure}[tb]{0.175\textwidth}
        \includegraphics[width=\textwidth]{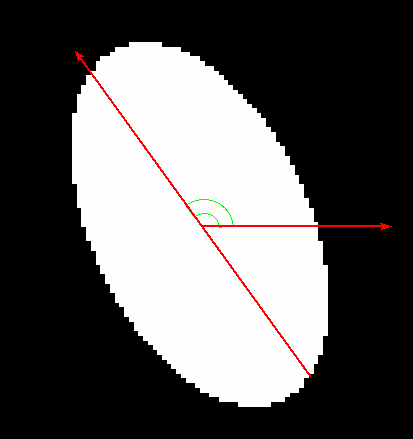}
        \caption{}
    \end{subfigure}\\
    \begin{subfigure}[tb]{0.175\textwidth}
        \includegraphics[width=\textwidth]{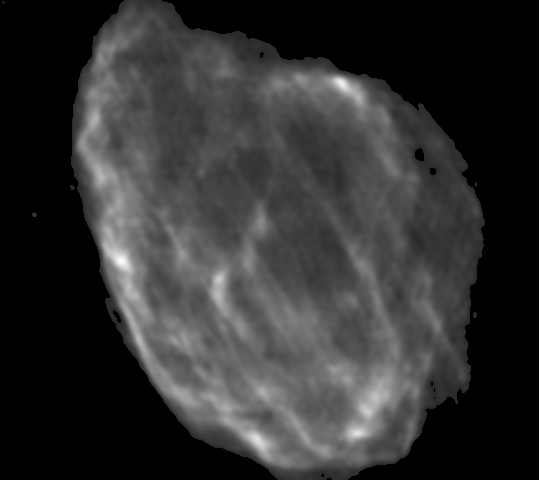}
        \caption{}
    \end{subfigure}
    \caption{Templates for the spatial modeling of SNR \w 44. a)-b) Elliptical ring (whole and divided) from \cite{w44_fermi_abdo_2010}. c)-d) Best full-ellipse (whole and divided). e) Radio template (1420MHz) from the THOR survey \citep{thor_survey_beuther_2016}.}
	\label{w44_templates}
\end{figure}

We considered the following spatial templates (Fig. ~\ref{w44_templates}) to describe the morphology of the remnant: 
\begin{itemize}
    \item the 4FGL catalog template, derived from \citet{w44_fermi_abdo_2010} after only two years of operations of \textit{Fermi}-LAT, consisting of an elliptical ring;
    \item a simple full-ellipse template;
    \item a radio (1420 MHz) template derived from The HI/OH/Recombination line survey of the Milky Way (THOR) 
    \citep{thor_survey_beuther_2016} \footnote{A more recent radio image was released in \citet{goedhart_1.3ghz_GPS_survey} where the authors performed a full polarization calibration for the pointing centered on the SNR \w44.}.
\end{itemize}
We derived the dimension and orientation of the full-ellipse template with a dedicated analysis, at first by fitting templates with varying inclination angles with 10$^{\circ}$ steps in the range [105$^{\circ}$, 145$^{\circ}$] and semimajor and semiminor axes with 0$\dotdeg$04 steps in the intervals [0.3$^{\circ}$, 0.46$^{\circ}$] and [0.16$^{\circ}$, 0.32$^{\circ}$]. The centroid of the ellipse was kept fixed at the coordinates taken from the 4FGL catalog template. We used the AIC to determine the best elliptical template. The minimum AIC value was obtained for an inclination angle of 115$^{\circ}$, a = 0$\dotdeg$38, b = 0$\dotdeg$2. A finer step of 0$\dotdeg$01 was used for the ellipse semiaxes by fixing the inclination angle to 115$^{\circ}$. The log-likelihood was maximized and the AIC minimized for [$a=0\dotdeg41\pm 0\dotdeg01$, $b=0\dotdeg23 \pm 0\dotdeg01$]. 
We also divided the catalog template and the full-ellipse template, similarly to what was done by \citet{Ajello_2016}, in search for the best division angle, starting from the major axis and considering another 18 angles in the range $\pm90^{\circ}$ around the central one. In Figs. \ref{w44_templates}b) and \ref{w44_templates}d) we show the divided templates. The best division angle for the elliptical ring is 115$^{\circ}$, and for the full ellipse, it is 125$^{\circ}$ with respect to the horizontal axis. We fit the two half-ellipses as independent sources (doubling the number of free parameters).

In Table \ref{results_first_templates} we report the results of the HE (1\,GeV -- 2\,TeV) \fermilat analysis obtained for the various models we adopted to describe the RoI. To calculate the number of degrees of freedom (d.o.f.), we considered 3 d.o.f. for each independent source employed to model \w44 (since the spatial component was fixed and the spectral model was an LP), 2 d.o.f. for each point-like and 3 d.o.f. for each extended source, 2 d.o.f. for each source with a PL spectral model, and 3 d.o.f. for each source with a LP spectral model. In this energy range, no source was characterized by a BPL spectral model.

The maximum value of log-likelihood and the minimum AIC value were obtained with a model that incorporates the radio template to describe the emission from \W44 itself and an additional disk template, hereinafter referred to as the diffuse disk, to describe centrally located diffuse gamma-ray emission in the surroundings of \W44 that is otherwise unaccounted for. Four sources were used to describe the region around the remnant in addition to the radio template and the diffuse disk: two moderately extended sources located at opposite edges along the major axis of the SNR, PS J1855.5+0141e and PS J1856.9+0102e, and two additional point-like sources detected by the source-finding algorithm, which correspond to the 4FGL sources 4FGL J1858.3+0209 and 4FGL J1859.3+0058. The extension of PS J1856.9+0102e also includes the locations of the two sources that were previously removed from the RoI, namely 4FGL J1857.4+0106 and 4FGL J1857.1+0056.
PS J1855.5+0141e and PS J1856.9+0102e correspond to the previously detected sources SRC-1 and SRC-2 in \cite{uchiyama_fermi-lat_2012}. We use the convention of the more recent \cite{peron_gamma-ray_2020}, who referred to them as NW and SE, respectively. As previously discussed, these are spatially modeled as disks. Table \ref{src1_src2_diff_disk} reports the best-fit position and radius of the diffuse disk and the NW and SE extended sources.

\begin{figure}[t]
	\centering
	\includegraphics[width=0.99\columnwidth, angle=0]{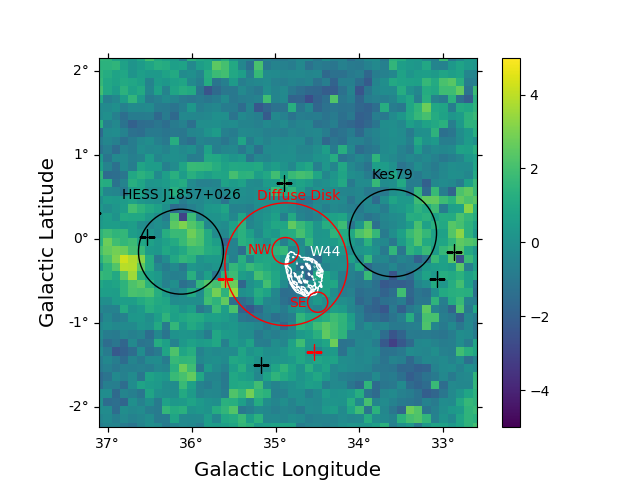}
	\caption{\textit{Fermi}-LAT deviation probability map, or PS map, of the \w 44 surroundings with the best model obtained from the likelihood analysis. The white contours represent the \W 44 radio template adopted in the analysis, derived from the template reported in Fig. \ref{w44_templates}e). The red crosses and circles show the new point-like and extended sources we added and fit in this analysis, and the black crosses and circles correspond to sources in the 4FGL-DR2 \textit{Fermi}-LAT catalog for which the morphology was not changed. The radii of the circles represent the $r_{68}$ of the extended sources that were modeled with a disk shape.}
	\label{fig:residmap_1GeV}
\end{figure}

Fig. \ref{fig:residmap_1GeV} shows a deviation probability map\footnote{The deviation probability map makes use of the p-value statistic (PS) data/model deviation estimator developed by \citet{psmap_bruel} and is sensitive to positive and negative fluctuations.}
that we obtained with the best-fit model.
Following the prescriptions in \cite{Abdollahi_2022}, we set the PS map parameters to $p_0=4^\circ$, $p_1=0.9$ and $p_2=0.1^\circ$. We also produced another PS map with parameters optimized for extended sources, that is, $p_0=5^\circ$, $p_1=0.8$, and $p_2=0.5^\circ$. No significant deviations (larger than $\pm 2\sigma$) were found in the central part of the RoI in this case either.

\begin{figure}[b]
	\centering
	{\includegraphics[width=\columnwidth, angle=0]{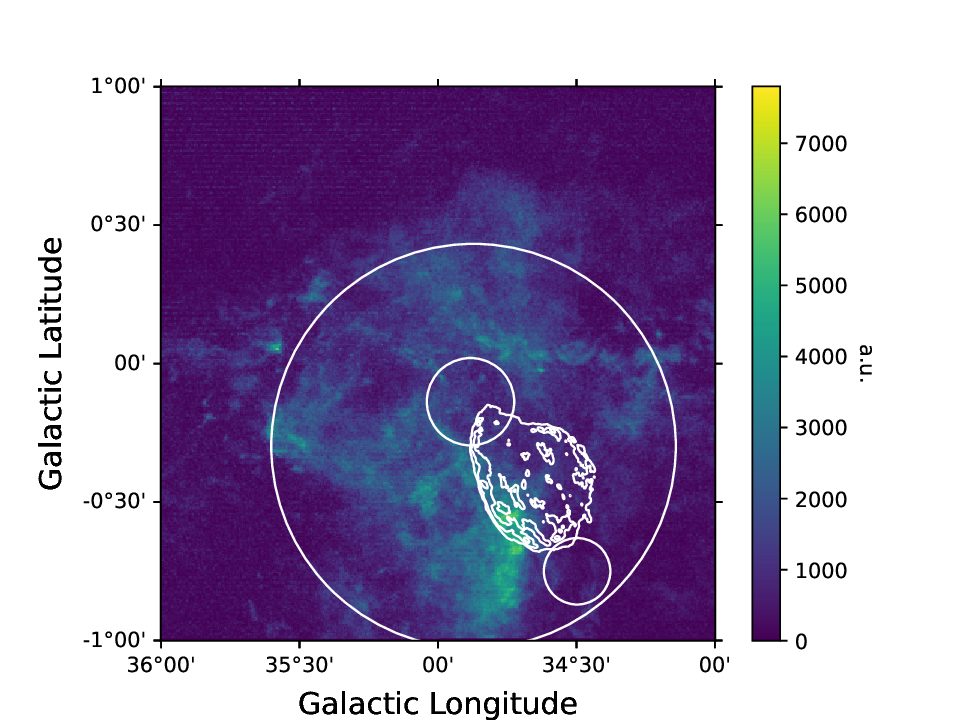}}
	\caption{CO template derived from FUGIN survey in the velocity interval (38.6, 49.8) km/s. The template is the sum of the $^{12}$CO and $^{13}$CO data. For comparison, the white contours and circles represent the W44 radio contours and the extended sources derived from the \textit{Fermi}-LAT analysis.}
	\label{fig:CO_templates_best}
\end{figure}

\begin{table*}
	\centering
    \caption{Results for 4FGL, full ellipse, and radio templates.}
	\begin{tabular}{lccccccccc}
        \toprule \toprule
		\textbf{Template (\W44)} & \multicolumn{1}{c}{\textbf{lnL}} & \multicolumn{1}{c}{\textbf{\#W44}} & \multicolumn{2}{c}{\textbf{Spatial}} & \multicolumn{2}{c}{\textbf{Spectral}} & \multicolumn{1}{c}{\textbf{k (d.o.f.)}} & \multicolumn{1}{c}{\textbf{AIC}} & \multicolumn{1}{c}{\textbf{$\Delta_{AIC}$}}  \\
         &  &  & \multicolumn{1}{c}{\#point-like} & \multicolumn{1}{c}{\#Extended} & \multicolumn{1}{c}{\#PL} & \multicolumn{1}{c}{\#LP} &  &  & \\
        \midrule
		Elliptical ring               & 57702     & 1     & 1   & 2   & 2  & 1  & 18     & -115368      & 290    \\
		Elliptical ring divided       & 57805     & 2     & 2   & 2   & 3  & 1  & 25     & -115560      & 98    \\
		Full ellipse                  & 57743     & 1     & 1   & 2   & 2  & 1  & 18     & -115450      & 208    \\
		Full ellipse divided          & 57804     & 2     & 2   & 2   & 3  & 1  & 25     & -115558      & 100    \\
		Radio (1420 MHz)              & 57856     & 1     & 2   & 3   & 4  & 1  & 27     & -115658      & 0    \\                       
        \bottomrule
	\end{tabular}
	\tablefoot{Between Cols. 3 and 7, we report the number of independent sources employed to model \w44, the number of point-like and extended sources, the number of sources with a PL or an LP spectral model. $\Delta_{AIC}$ was calculated with respect to the minimum AIC, which corresponds to the radio template. }
	\label{results_first_templates}
\end{table*}
\begin{table}
	\centering
    \caption{Fit position and radius of the three extended sources of Fig. \ref{fig:residmap_1GeV}.}
	\begin{tabular}{lccc}
        \toprule
  	\textbf{Sources} & \textbf{$l$ (deg)} & \textbf{$b$ (deg)}
        & \textbf{$r_{68}$ (deg)}  \\
  
        \midrule
        NW        & 34.88$\pm$0.01            & -0.14$\pm$0.01              & 0.16$\pm 0.01$                   \\
        SE        & 34.49$\pm$0.01            & -0.75$\pm$0.01               & 0.12$\pm 0.01$                   \\
        Diffuse Disk          & 34.87$\pm$0.02            & -0.30$\pm$0.02               & 0.73$\pm 0.02$   \\
        \bottomrule
	\end{tabular}
	\tablefoot{$l$ and $b$ are the galactic coordinates, and $r_{68}$ represents the 68\% containment radius of the best-fit spatial model.}
	\label{src1_src2_diff_disk}
\end{table}

\subsection{Diffuse disk morphology}
\label{sec:diff_disk_morph_analysis}
We also investigated the morphology of the diffuse emission in the surroundings of \w44 in detail. In order to find possible associations with gas complexes, we studied $^{12}$CO and $^{13}$CO data from the FOREST Unbiased Galactic plane Imaging survey with the Nobeyama 45-m telescope (FUGIN)\citep{Umemoto_2017} in a region within $\sim$ 1$^{\circ}$ from the SNR.
We chose nine velocity intervals between 30\,km/s and 90\,km/s from the emission profiles of the gas in correspondence of the large disk and in correspondence of the NW and SE disks. The velocity intervals corresponded to peaks of the gas profile within each of these regions.
For each case, we derived a template and repeated the likelihood-fitting procedure for \textit{Fermi}-LAT data. To do this, we adopted the radio template for the \W 44.

We refit the other sources in the RoI, including the NW and SE sources, spatially and spectrally in each case. We used the AIC value to decide which was the best configuration. 
Of the CO templates we derived, the template that provided the best results corresponded to the velocity interval (38.6 - 49.8)\,km/s, which is shown in Fig. ~\ref{fig:CO_templates_best}. However, the large disk derived previously (and reported in Table \ref{src1_src2_diff_disk}) still provided the best description of the RoI, with a $\Delta$AIC=10.6 in favor of the disk hypothesis. Therefore, we also used this latter model as reference spatial model for the MAGIC analysis.

When the CO template is used instead of the large disk to model the large diffuse background, the spatial and spectral parameters of the NW and SE sources did not change significantly. This is compatible with the results in Table \ref{src1_src2_diff_disk} within the statistical uncertainties.

\subsection{Spectral analysis}
\label{sec:spect_analysis}

For the low-energy spectral analysis (E~$>$~100\,MeV), we followed the weighted likelihood approach. This choice was driven by the necessity of accounting for systematic errors, whose main origin is the diffuse background emission. In particular, the Galactic interstellar emission model is affected by systematic uncertainties, which shows imperfections that stem from assumptions made in its construction and uncertainties in the adopted templates. This is a crucial point in particular for the source analysis we performed over energies below a few hundred MeV, as in the case of this work. In this regime, where the PSF is several degrees wide, the source-to-background ratio is low (at the percent level), and the systematic errors on the background model are critical. 
We carried out the weighted likelihood analysis following the procedure reported in \cite{Ballet_2015} and in Appendix B of \cite{abdollahi_fermi_2020}.

\begin{figure}[htb]
	\centering
	{\includegraphics[width=0.99\columnwidth, angle=0]
    {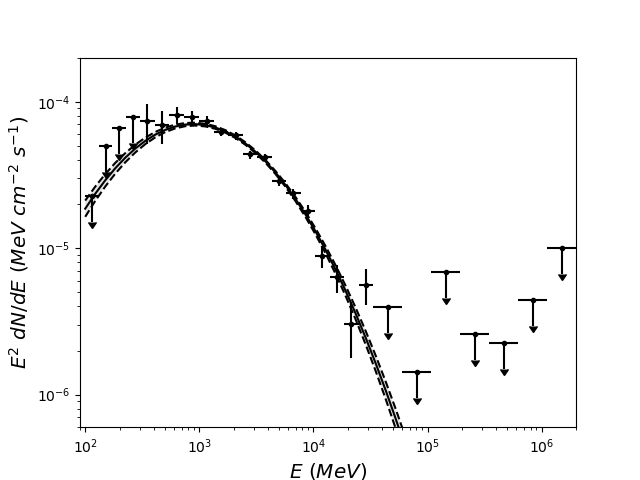}}
	\caption{SED of SNR \w 44 (only \fermi-LAT data). The solid line shows the best-fit curve obtained with an LP model, and the dashed lines represent the 1 $\sigma$ uncertainty.}
	\label{fig:w44_sed}
\end{figure}

\begin{figure}[htb]
	\centering
	{\includegraphics[width=0.99\columnwidth, angle=0]
	{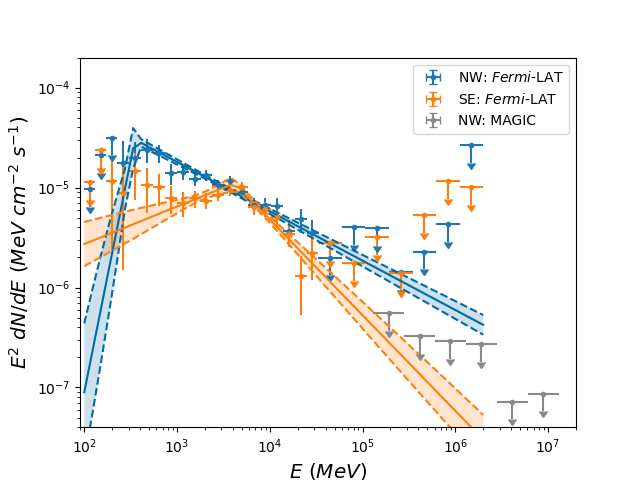}}
    \caption{SEDs of the two sources NW and SE with \fermi-LAT data and MAGIC ULs.}
	\label{fig:seds_NW_SE}
\end{figure}

We performed this analysis by fixing the spatial models for the sources of interest to those obtained from the high-energy analysis described above. Since for this spectral analysis we considered a wider energy range, we repeated the curvature fitting procedure on the newly added sources within 1$^{\circ}$ from the center of the RoI, and we reevaluated the $TS_{curv}$ for each of them. We also repeated the fit of the spectral parameters of the two extended sources Kes79 and HESS J1857+026. The best spectral model for the diffuse disk is the LP, as shown in Table \ref{tab:ts_curv}, while BPL is the best model for the NW and SE sources. In Tables \ref{tab:spectral_params_LP} and \ref{tab:spectral_params_BPL} we report the best-fit spectral parameters for the \w44 SNR, which has an LP spectral model, and for the three close sources. We evaluated the systematic errors from the uncertainties of the effective area of the instrument.
The resulting spectral energy distribution (SED) for \w44 is shown in Fig. \ref{fig:w44_sed} and the SEDs related to the sources NW and SE are displayed in Fig. \ref{fig:seds_NW_SE}. We added the systematic errors in quadrature to the errors of the respective source.

In order to probe the presence of a bump in the spectra of the sources NW and SE, which might arise at energies where the spectrum of the \w44 SNR starts to decrease, we performed an analysis over a shorter energy range, 300 MeV - 30 GeV, and repeated the curvature test, now focused on how well a BPL model can describe the spectra of sources \w44, NW, and SE. Fitting a BPL spectral model to the spectrum of the \w44 SNR, we obtained a significance $TS_{curv}=397.1$ and a value of $E_{br}=(1.7\pm 0.1)\ GeV$. For source NW, the BPL significance is $TS_{curv}=8.6$ with $E_{br} = (0.47\pm0.07)\ GeV$, and for source SE, the BPL significance is $TS_{curv}=22.1$ with $E_{br}=(5.4\pm1.2)\ GeV$. 

\begin{table}
    \centering
    \caption{Values of $TS_{\rm curv}$ obtained with LP and BPL as spectral templates.}
	\begin{tabular}{lccc}
        \toprule
		\textbf{Sources} & \textbf{$TS_{\rm curv/LP}$} & \textbf{$TS_{\rm curv/BPL}$} \\
        \midrule
        NW          & 13.3                  & 24.8      \\
        SE          & 23.6                  & 34.9      \\
        Diffuse Disk     & 173.1            & 140.7     \\
        \bottomrule
	\end{tabular}
	\label{tab:ts_curv}
\end{table}

\begin{table*}
	\centering
    \caption{Best-fit spectral parameters of the sources with LP as spectral model, obtained from the \textit{Fermi}-LAT spectral analysis. $E_b$ is a scale parameter that was fixed.}
	\begin{tabular}{lcc}
        \toprule \toprule
		\textbf{Sources} & \textbf{$N_0$} ($10^{-11}$ MeV$^{-1}$ cm$^{-2}$ s$^{-1}$) &  \textbf{$E_b$} (GeV)  \\
        \midrule
		\w 44            & $(0.99\pm0.02 (stat) _{-0.04} ^{+0.02} (syst))$  &  2.34  \\
		Diffuse Disk     & $(3.88\pm0.30 (stat) ^{+0.13}_{-0.10} (syst))$   &   1  \\
        \midrule \midrule
                    & \textbf{$\alpha$} & \textbf{$\beta$}  \\
        \midrule
        \w 44            & $2.54\pm0.03 (stat) \pm0.01 (syst)$  &  $0.28\pm0.02(stat) \pm 0.01 (syst)$  \\
		Diffuse Disk     & $1.87\pm0.12 (stat) \pm0.01 (syst)$  &  $0.34\pm0.06 (stat) \pm0.01 (syst)$ \\
        \bottomrule
	\end{tabular}
	\label{tab:spectral_params_LP}
\end{table*}

\begin{table*}
    \centering
    \caption{Best-fit spectral parameters of the sources with BPL as spectral model, obtained from the \textit{Fermi}-LAT spectral analysis.}
	\begin{tabular}{lcc}
    \toprule \toprule
    \textbf{Sources} & \textbf{$N_0$} ($10^{-12}$ MeV$^{-1}$ cm$^{-2}$ s$^{-1}$) & \textbf{E$_{br}$} (GeV) \\
    \midrule
	NW      & $(240\pm78(stat)\pm1 (syst))$  & $0.354\pm0.043(stat)\pm0.002 (syst)$    \\
    SE      & $(0.64\pm0.28(stat)\pm0.02(syst))$  & $(4.17\pm0.69(stat)\pm0.03(syst))$   \\
    \midrule \midrule
            & \textbf{$\gamma_1$} & \textbf{$\gamma_2$}  \\
    \midrule
    NW      & $2.69 \pm 3.75 (stat) ^{+0.09} _{-0.07} (syst)$ & -$2.49\pm0.05(stat)\pm0.01 (syst)$   \\
    SE      & -$1.62\pm0.25(stat)\pm0.01(syst)$ & -$2.96\pm0.13(stat)\pm0.01(syst)$   \\
    \bottomrule
	\end{tabular}
	\label{tab:spectral_params_BPL}
\end{table*}

\subsection{Contribution of the Galactic diffuse component to sources NW and SE and to the diffuse disk}
\label{sec:galdiff_analysis}

We evaluated the contribution of the Galactic diffuse background from the model adopted for the \fermilat analysis (gll\_iem\_v07) inside the three circular regions identified by sources NW and SE and the diffuse disk. 
The resulting plots are reported in the Appendix \ref{sec:appendix}. 
In the circular region corresponding to the diffuse disk model (\ref{fig:galdiff_seds}.3), the Galactic diffuse background dominates all other contributions to the emission, including the diffuse disk source itself. By contrast, the NW and SE sources are larger than the Galactic diffuse background in their corresponding regions.
The diffuse disk spectrum is characterized by a cutoff similar to that of the Galactic diffuse model. This suggests that the diffuse disk emission, which corresponds to roughly 32\% of the originally fit Galactic diffuse flux, could also be produced by interaction of the Galactic CR sea with gas in the region.
However, we do not rule out other possible origins of this emission.

When we assume the same percentage ($\sim 32\%$) as the contamination of the Galactic diffuse background in the NW and SE regions, we obtain that only 10\% (20\%) of the total emission of NW (SE) can be explained, suggesting that most of the emission has a different origin. 
A more comprehensive discussion is presented in Sect. \ref{sec:model}.

\section{MAGIC observations and data analysis}
\label{sec:obs_magic}

The \gls{MAGIC} telescopes are a system of two \cglspl{IACT} located on the Canary island of La Palma at an altitude of 2200\,m above sea level \citep{aleksic_magic_upgradeI_2016}.
\gls{MAGIC} observed the \w44 region between April 2013 and August 2014 for 173.7\,h after quality cuts at zenith angles between $25\degr$--$45\degr$. The \gls{MAGIC} observations were centered on the coordinates of the NW source from \citet{uchiyama_fermi-lat_2012} using a wobble distance of $0\fdg4$. The low-level analysis was performed with the \gls{MARS} \citep{zanin_mars_2013,aleksic_magic_upgradeII_2016}. We used the spatial likelihood analysis package \texttt{SkyPrism} \citep{vovk_spatial_2018} for our subsequent high-level analysis.
An estimation of the systematic uncertainties of the MAGIC telescopes can be found in \citet{aleksic_magic_upgradeII_2016} and \citet{vovk_spatial_2018}. The uncertainties were directly calculated from the log-likelihood deviation from the obtained best fit.

\begin{figure}[thb]
  \centering
  \includegraphics[width=0.95\columnwidth,angle=0]{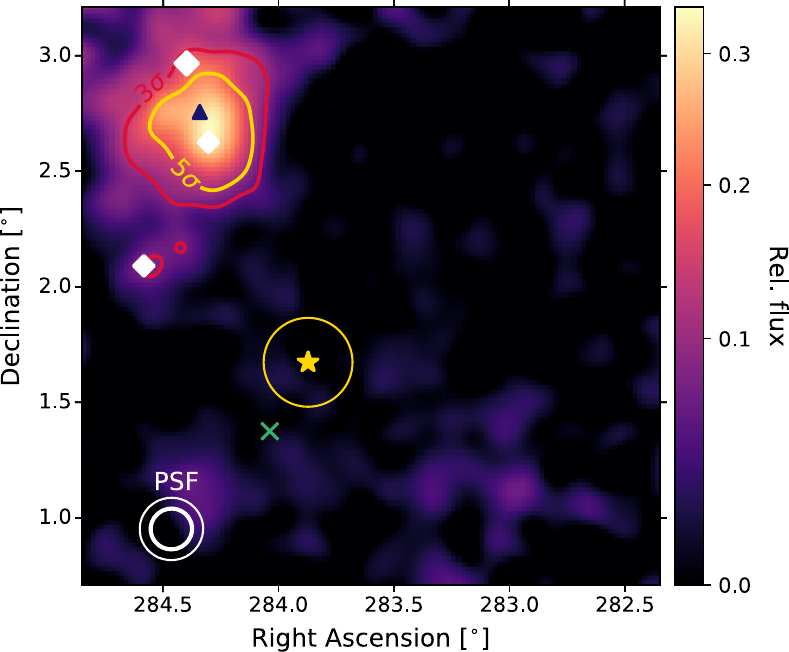}
  \caption{Relative flux map of the region observed with the MAGIC telescopes. The modeled sources MAGIC\,J1857.2+0263, MAGIC\,J1857.6+0297, and HESS\,J1858+020 are marked with white diamonds. The center of \w44 is marked with a green cross. We also show the location of HESS\,J1857+026 from \citet{abdalla_hess_2018} with a dark blue triangle. The NW center position is marked with a yellow star, and the extension is indicated by the yellow circle. The 39\% and 68\% containment contours of the MAGIC PSF are depicted in the bottom left corner.}
  \label{fig:w44_magic_RelFluxmap}
\end{figure}

From the wobble observations, we constructed the background camera exposure model using an exclusion map. We excluded the known sources in our field of view: namely, HESS\,J1857+026 \citep{abdalla_hess_2018} using an exclusion zone radius of $0\fdg45$, which is large enough to enclose the two known high-energy sources MAGIC\,J1857.2+0263 (RA: 18\h57\m13\farcs0, Dec: 02\degr 37\arcmin 31\arcsec) and MAGIC\,J1857.6+0297 (RA: 18\h57\m35\farcs6, Dec: 02\degr 58\arcmin 02\arcsec) \citep{aleksic_magic_2014}. We also excluded HESS\,J1858+020 \citep{abdalla_hess_2018} with a radius of $0\fdg17$, NW, and SE with their position and extension from the \fermilat analysis above. The centers of the exclusion regions around HESS\,J1857+026 and HESS\,J1858+020 were set to the respective maximum in our sky maps.
Due to the curved spectra of \w44 and the large background components, these sources did not need to be included in the \gls{MAGIC} energy range.
The relative flux map is presented in Fig. \ref{fig:w44_magic_RelFluxmap}.
No significant detection was obtained with \gls{MAGIC}. Therefore, we calculated $95\%$ CL upper limits (ULs) for the NW source in each of our energy bins (Fig. \ref{fig:seds_NW_SE}) following the method described by \citet{rolke_limits_2005}. The lowest-energy UL was lower than 3\% of the residual background, and we therefore fixed it at this higher level. This is equivalent, for instance, to the analysis of the gravitational wave event GW151226 in \citet{berti_study_2018}.
We did not extract ULs for \w44 because they would be more than an order of magnitude above the model curve. The SE source is smaller in radius than NW and closer to the edge of the MAGIC field of view. This means that if we were to estimate ULs for this area, they would be even higher than those calculated for the NW region. Therefore, we did not extract ULs for SE to avoid complicating the likelihood fitting unnecessarily.

\section{Modeling the gamma-ray emission}
\label{sec:model}

In this section, we show that the gamma-ray emission from W44 and its surrounding region can be understood in the framework of particle acceleration and escape from the forward shock.

We note that the NW and SE sources are both located along the major axis of W44, suggesting that the emission from these two sources might be due to accelerated CRs that escape from the W44 in the direction of the local magnetic field and that interact with the interstellar gas. This interpretation is supported by a recent analysis from \cite{Liu-Hu-Lazarian:2022}, who used the velocity gradient technique (VGT) to infer the direction of the large-scale magnetic field\footnote{The VGT is a technique based on the properties of magnetohydrodynamic (MHD) turbulence. Due to the reconnection, MHD turbulent eddies are stretched along local magnetic fields that percolate the eddies and thus become anisotropic \cite{Lazarian-Vishniac:1999}. The resulting anisotropy induces gradients of velocity fluctuations’ amplitude to be perpendicular to the directions of local magnetic fields, which enables the magnetic field tracing employing the amplitude of the turbulence’s velocity fluctuations that are available from observations.}. 
The VGT was applied to formyl ion (HCO$^+$), atomic hydrogen (HI), and CO emission lines from MCs that interact with \W44, showing that the local magnetic field is oriented along the major axis of the SNR. This result also agrees with radio polarization measurements in the SE region \citep{goedhart_1.3ghz_GPS_survey} and with the Plank data at 353 GHz, which measures the dust polarization, at least for regions with high-intensity emission \cite[see discussion in][sec.~4.2]{Liu-Hu-Lazarian:2022}. 
Hence, the absence of gamma-ray emission from other locations around the SNR could just be the consequence of anisotropic escape of particles in the direction of the local magnetic field, as reported by \cite{peron_gamma-ray_2020}. A similar scenario, originally proposed by \cite{Gabici+:2009}, was applied to W28 \citep{Nava-Gabici:2013}, even though in their case, there was no indication about the direction of the external magnetic field. 

\begin{figure}[t]
	{\includegraphics[width=0.99\columnwidth]{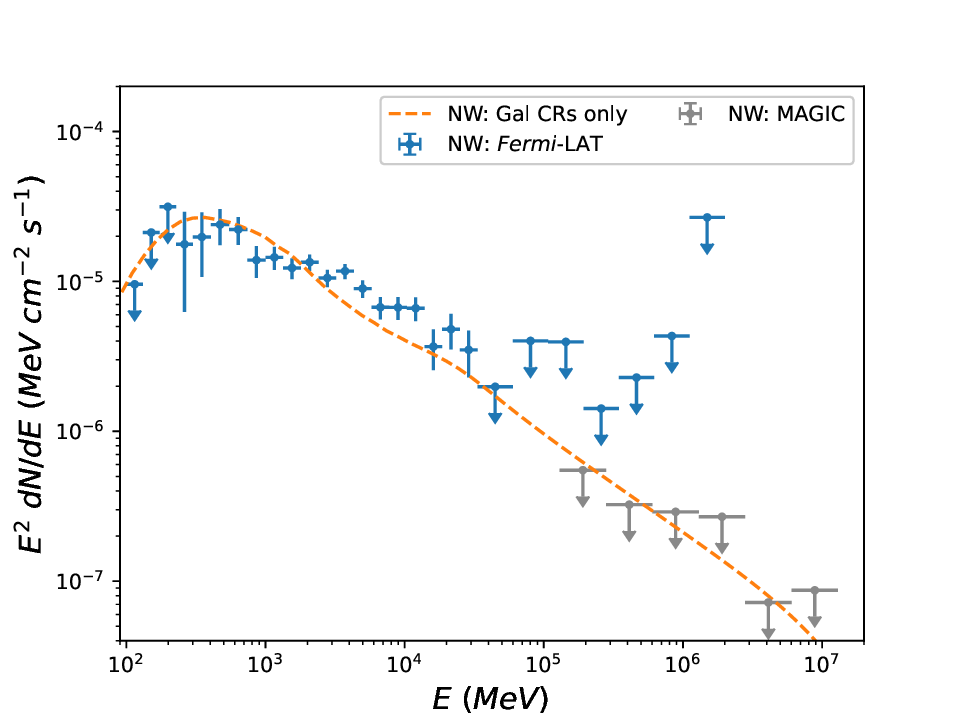}}
	\caption{Gamma-ray emission from source NW compared to the pp emission due to Galactic CRs from a target mass equal to $5.5 \times 10^4\, M_{\odot}$.}
	\label{fig:SRC1-GalCR}
\end{figure}

Alternatively, the emission from the two sources might result from the interactions of the Galactic CR sea with target gas that is not accounted for in the model for the diffuse emission \citep{Acero_2016}. The mass estimated in the diffuse model is affected by an uncertainty of 30\% mainly due to the CO-to-H$_2$ conversion factor \citep{Bolatto+2013}.  To check this scenario, we estimated the mass of the gas needed to produce a flux compatible with the \fermilat data from NW source assuming that the Galactic CR spectrum is equal to the local spectrum measured by the Alpha Magnetic Spectrometer (AMS)-02 \citep{AMS02-proton:2015}. The corresponding plot is shown in Fig. ~\ref{fig:SRC1-GalCR}, where we also report MAGIC upper limits. The amount of mass required to fit \fermilat data is $\sim 5.5\times 10^4 \, M_{\odot}$. This value corresponds to $\sim 30\%$ of the total mass included in the diffuse model in the NW region integrated along the line of sight. Hence, in principle, the additional mass would be compatible with the mass uncertainty in the diffuse model. The total gamma-ray emission integrated along the line of sight also depends on the mass location, however. The total NW source emission can only be explained when the entire mass is concentrated at the NW source location, resulting in a gamma-ray emission that is three times higher than is predicted from the diffuse model (see Appendix \ref{sec:appendix}). In addition, the predicted spectrum does not reproduce the harder \fermilat spectrum at $\sim 10$\,GeV and slightly overshoots the MAGIC upper limits. 

In order to estimate the goodness of this model, we fit the overall flux of the predicted model by keeping the spectral shape fixed. We first only used \fermilat data and then added the MAGIC spectral points. In the latter case, we only used LAT data points up to 100 GeV because the MAGIC upper limits are more constraining for energies above 100 GeV.
We calculated the $\chi^2$ value of the best-fit models in the two cases and report them in Table \ref{chi2_galcr} together with the corresponding p-values. We concluded that the model can be rejected at the $\sim3.6\sigma$ confidence level using \fermilat data alone and at the $\sim5.8 \sigma$ confidence level with \fermilat + MAGIC data.
For this reason, a scenario in which the gamma-ray emission is due to interactions with the Galactic CR sea is disfavored. In the next section, we focus on the escape scenario of freshly accelerated CRs.

\begin{table}
    \centering
    \caption{Goodness of fit of the model reported in Fig.\ref{fig:SRC1-GalCR}}
	\begin{tabular}{lcc}
        \toprule
		\textbf{} & \textbf{$\chi^2 / N$} & \textbf{$p$-value}  \\
        \midrule
		\fermilat only        & 61.77/27  & $1.5 \cdot 10^{-4}$  \\
		\textit{Fermi}-LAT+MAGIC       & 95.09/28  & $3.1 \cdot 10^{-9}$  \\
        \bottomrule
	\end{tabular}
	\tablefoot{Chi-squared values $\chi^2$, number of data points $N$ and corresponding $p$ values obtained with the Galactic CR emission model. The values were calculated using the \fermilat data alone and \fermilat + MAGIC data.}
	\label{chi2_galcr}
\end{table}

\subsection{Model for particle acceleration and escape}
We adopted the model developed in \cite{Celli:2019} that was applied to $\gamma$-Cygni \citep{MAGIC_gamma-Cygni:2023} and to the Cygnus Loop \citep{Loru:2021}. This model assumes spherical symmetry for acceleration and escape. However, as already noted, we wished to account for particles that only escape along the magnetic flux tube oriented SE-NW. To first approximation, this scenario can be described by our spherical model assuming that all external particles located at distance $r$ from the SNR forward shock are located at the same distance, but in a cylinder with a radius equal to the shock radius.  The following equations and assumptions are based on the model described in \cite{Celli:2019}.

For the SNR properties, we adopted the values that are most commonly used in the literature. The SN explosion kinetic energy was fixed to $E_{\rm SN}=10^{51}$~erg, the circumstellar density was $n_0= 10 \, \rm cm^{-3}$, and the SNR age was $t_{\rm age}= 20$\, kyr \cite[see the SNRcat\footnote{\url{http://snrcat.physics.umanitoba.ca/}} from][]{Ferrand-Safi-Harb:2012}. The distance was estimated in the range 2.1-3.3 kpc; we assumed $d= 2.2$\, kpc following \cite{peron_gamma-ray_2020}. We then followed \cite{Truelove_McKee99} to describe the SNR evolution assuming an ejecta power-law index equal to $0$. The ejecta mass was fixed to $5\,M_{\odot}$ to match the average shock radius of $0\dotdeg29$, which corresponds to $R_{\rm sh}\simeq 11$~pc at a distance of $2.2$~kpc.

We accounted for particle acceleration at the forward shock assuming that a fixed fraction of the shock kinetic energy is transferred to nonthermal particles. The instantaneous proton spectrum accelerated at the shock is
    \begin{equation} 
    \label{eq:f_0}
     f_{p,0}(p,t) = \frac{3 \, \xi_{\rm CR} u^2_{\rm sh}(t) \rho_0}{4 \pi \, c (m_p c)^4  \Lambda(p_{\max}(t))} \left( \frac{p}{m_p c} \right)^{-s}  e^{ -p/p_{\max}(t)} \,,
    \end{equation}
where $\rho_0=n_0 m_p$ ($m_p$ is the proton mass), $\Lambda(p_{\max}(t))$ is a normalization constant such that the CR pressure at the shock is $P_{\rm CR} = \xi_{\rm CR} \rho_0 u_{\rm sh}^2$, $u_{\rm sh}$ being the shock velocity. The factor $\xi_{\rm CR}$ represents the instantaneous acceleration efficiency and was kept constant during the whole evolution of the SNR. The spectral slope $s$ was taken as free parameter and expected to be close to 4.
$p_{\max}(t)$ represents the instantaneous maximum momentum (and $E_{\max}$ is the corresponding maximum energy) achieved at time $t$. Simple considerations on particle confinement suggest that $p_{\max}$ decreases during the Sedov-Taylor (ST) phase \citep{Gabici+:2009}, and for simplicity, it was assumed to follow a simple power law, 
    \begin{equation} 
    \label{eq:pmax0}
     p_{\max}(t) =
      \begin{cases} 
       p_\textrm{M} \left( t/t_{\rm ST} \right)     & \text{if } t \leqslant  t_{\rm ST} 	\\
       p_\textrm{M} \left( t/t_{\rm ST} \right)^{-\delta}     & \text{if } t > t_{\rm ST}
      \end{cases},
    \end{equation}
%
where $t_{\rm ST}\simeq 420$\,yr is the beginning of the ST age. The absolute maximum momentum $p_{\rm M}$ and the slope $\delta$ are free parameters that are constrained from the data. The maximum energy was assumed to increase linearly during the ejecta-dominated phase ($t<t_{\rm ST}$). However, the final spectrum at the present age $\left(t_{\rm age}\right)$ of the SNR is not very sensitive to the initial phase of the remnant evolution because $t_{\rm ST} \ll t_{\rm age}$. Hence, the behavior of the maximum energy during the ejecta-dominated phase (i.e. for $t<t_{\rm ST}$) cannot be constrained. 

Particles with momentum $p>p_{\max}(t)$ escape from the shock and start to diffuse with a diffusion coefficient that we assumed to be proportional to the Galactic one, that is, $D_{\rm ext}= \chi D_{\rm gal}$, where $D_{\rm gal}(p) = 3.6 \times 10^{28} \beta\, (p\, c/\rm GeV)^{1/3}\, cm^2 s^{-1}$. 

\begin{figure}[t]
	{\includegraphics[width=0.99 \columnwidth]{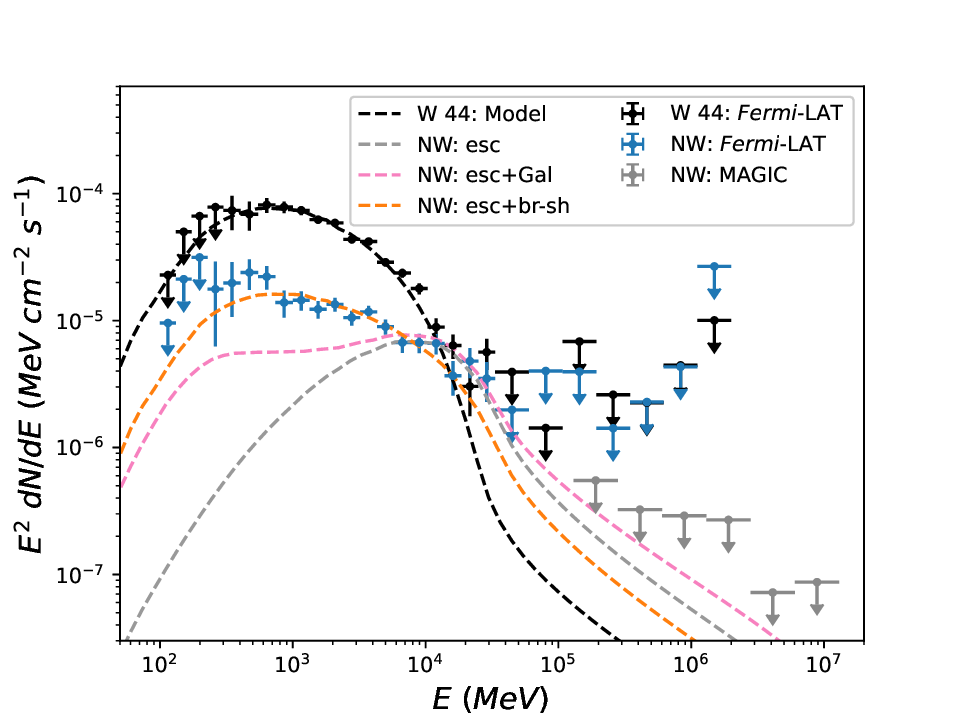}}
	{\includegraphics[width=0.99 \columnwidth]{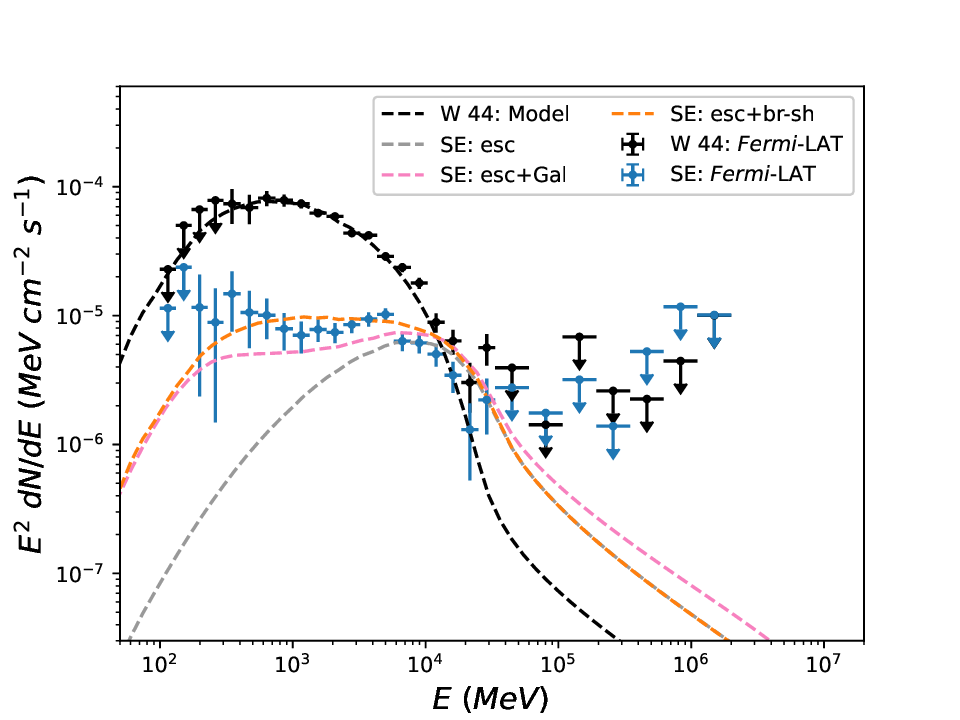}}
    \caption{Gamma-ray emission from sources NW (top panel) and SE (bottom) compared to our model. The dashed gray line shows the emission from the high-energy particles escaping from the shock alone. The dashed pink line also includes an additional component due to Galactic CRs interacting with the cloud material, and the dashed orange line shows the sum of the dashed gray line with the contribution from the low-energy particles escaping from the broken-shock region. For comparison, the black points and dashed lines refer to the \w44 emission.}
	\label{fig:broken-shock}
\end{figure}

In summary, the model had seven parameters, $\xi_{\rm CR}$, $s$, $p_{\rm M}$, $\delta$, and $\chi$, plus the two masses of the clouds, $M_{\rm NW}$ and $M_{\rm SE}$, that were constrained using the gamma-ray data.
The gamma-ray emission from proton-proton interaction was calculated using the parameterization provided by \cite{Kafexhiu-Taylor2014}. The emission has two components: the first one comes from the interior of the SNR, while the second one comes from the exterior and diffuses mainly along the magnetic flux tube. 
Fig. ~\ref{fig:broken-shock} shows the result of our model compared to data from  \w44, NW and SE. The emission from \w44 is well fit using $\xi_{\rm CR}= 1.3\%$, $s= 4.2$, $p_M= 100$\, TeV and $\delta= 2$.
As noted by other authors \citep{Ackermann_2013} and as we also found in Sect. \ref{sec:spect_analysis}, the spectrum from \w44 has a break at a few GeV. In our scenario, this break corresponds to protons with an energy equal to $E_{\rm max}(t_{\rm age}) = 44$\,GeV. Above this energy, the spectrum steepens because particles escape from the remnant interior. 

It is quite remarkable that the emission from the SE source shows a bump at roughly the same energy at which \w44 shows a break, suggesting that the peak is due to escaping particles with $p \gtrsim p_{\max}(t_{\rm age})$. The NW source instead does not show clear evidence of a similar bump, which might have been smoothed due to the broken-shock scenario explained in Sect. \ref{sec:model_broken_shock}, however. To reproduce the observed flux at a few GeV, we needed to assume cloud masses of $M_{\rm NW}= 9.9\times 10^3 M_{\odot}$ and $M_{\rm SE}= 8.7 \times 10^3 M_{\odot}$ and a diffusion coefficient around the SNR that is suppressed by a factor of 5 compared to the average Galactic diffusion coefficient, namely $\chi= 0.2$. For higher values of $D_{\rm ext}$, the gamma-ray flux above $\sim 4$\,GeV from the W44 interior and the two clouds would be suppressed because particles would escape faster from the region.
However, the predicted gamma-ray spectra below $\sim 4$\,GeV are suppressed because particles with $pc\lesssim 44$\,GeV are still confined inside the SNR. The escape model in this basic version is therefore unable to account for the emission from the NW and SE clouds below a few GeV. We tried to include an additional component due to preexisting Galactic CRs (dashed pink lines in Fig. ~\ref{fig:broken-shock}) that interact with the cloud masses. In principle, this component is already accounted for in the background model. However, if the total mass along the line of sight is underestimated, a residual component may still be present. By adding this component, the discrepancy with the data is reduced (especially for SE), but is still unsatisfactory.

Before we proceed, we stress that in the model we proposed, the gamma-ray emission is dominated by freshly accelerated particles, while other works invoked the reacceleration of preexisting CRs \citep{uchiyama_fermi-lat_2012, Cardillo+2016}. The main difference between these two approaches is connected to the average density of the circumstellar medium in which the shock expands. For a density $\gtrsim 100 \, \rm cm^{-3}$, the shock becomes radiative, and the combination of reacceleration plus compression is enough to account for the gamma-ray emission. Conversely, we assumed a much lower density,  $n_0 \simeq 10\, \rm cm^{-3}$, and relied on the results obtained by \cite{peron_gamma-ray_2020} through the analysis of the CO emission. It is known that the circumstellar medium around W44 contains some denser clumps that interact with the shock, and these easily enter the radiative phase \citep{Cosentino+2018}. It is currently unclear which the filling factors of these regions are, however.

\subsection{The broken-shock scenario} 
\label{sec:model_broken_shock}
From the previous section, it is clear that the escape model fails to explain the lowest energies from the external sources, especially for the NW source, where the detected emission is higher by one order of magnitude at least than the model prediction. For this reason, we propose a novel scenario in which some number of particles with an energy lower than $E_{\rm max}(t_{\rm age})$ is allowed to escape from the system. If the shock were a continuous surface, the escape of low-energy particles would not be possible, but in the case of middle-aged SNRs that expand in highly inhomogeneous media, the shock surface could be broken where the upstream medium is denser, resulting in the escape of a small fraction $f_{\rm br}$ of the total accelerated particles. As a consequence, even low-energy particles might fill the region outside of the shock up to a distance equal to their diffusion length, $l_{\rm diff} = \sqrt{2 D_{\rm ext} t}$. If the emitting region has a size comparable to or smaller than $l_{\rm diff}$, we may expect gamma-ray emission also at an energy lower than $\sim 0.1 E_{\rm br}$.
We refer to this model as the broken-shock scenario.

The precise modeling of this scenario is beyond the scope of the present work because it depends on quantities that are not known with sufficient accuracy. In particular, we need to know the density distribution of the circumstellar medium to describe the time evolution of the shock propagating in such a medium.

We limit our discussion here instead to evaluate whether the broken-shock scenario is compatible with the model presented in the previous section in terms of the number of escaped particles, diffusion length, and maximum energy. In particular, concerning the latter, we verified that the fragmentation of the shock surface does not reduce the maximum energy that can be achieved in the system. In other words, the shock surface should be continuous for a typical size $l_{\rm sh}$ larger than the upstream diffusion length, $D(p)/u_{\rm sh}$, of particles with a momentum equal to $p_{\max} = 44 \, \rm GeV/c$, as estimated in the previous section. We assumed that the diffusion in the shock region is due to magnetic turbulence that is produced through the resonant streaming instability, such that the diffusion coefficient can be written as $D= r_L c/(3 \mathcal{F})$, where $r_L(p)$ is the Larmor radius, and $\mathcal{F}$ is the power of the magnetic turbulence for a logarithmic bandwidth. In quasi-linear theory, this latter quantity is given by \cite[see, e.g.][]{Blasi:2013}
\begin{equation}
   \mathcal{F} = \frac{\pi}{2} \frac{\xi_{\rm CR}}{\ln{(p_{\max}/m_p c)}} \frac{u_{\rm sh}}{v_{\rm A}} ,
\end{equation}
where $v_A = B_0/(4\pi n_0 m_p c)^{1/2}$ is the Alfv\'en speed upstream of the shock, and $B_0$ is the large-scale magnetic field. Hence, the condition $l_{\rm sh} > D(p)/u_{\rm sh}$ reads
\begin{equation}
    l_{\rm sh} > \frac{2}{3\pi} \frac{\ln{(p_{\max}/m_p c)}}{\xi_{\rm CR}} \frac{r_L(p_{\max}) c v_A}{u_{\rm sh}^2} \simeq 0.01 \, \rm pc,
\end{equation}
where the last equality was obtained using $n_0 = 10\, \rm cm^{-3}$. Interestingly enough, this result does not depend on the value of $B_0$. The lower limit for the shock size is very low and should not be a limiting condition. Detailed radio maps suggest that $l_{\rm sh}$ is even much larger \citep{Cotton_2024, goedhart_1.3ghz_GPS_survey}.

We made the simplifying  assumption that in addition to the escape of high-energy particles with $E>E_{\rm max}(t_{\rm age})$, a fraction $f_{\rm br}$ of low-energy particles (i.e., $E< E_{\rm max}(t_{\rm age})$)  also escapes from the SNR interior and propagates in the circumstellar medium according to the diffusion coefficient $D_{\rm ext}$ estimated in the previous section. The value of $f_{\rm br}$ was then determined from the comparison with the gamma-ray data.
The dashed orange lines in Fig. ~\ref{fig:broken-shock} show the total gamma-ray emission resulting from the broken-shock scenario assuming a value $f_{\rm br}= 0.2$ and $0.1$ for NW and SE, respectively.
Because a relatively small fraction of 10-20\% allows us to increase the expected flux at low energies to levels similar to the measured ones, this option can be considered a possible explanation for the observed flux.
The estimated value of $f_{\rm br}$ is low enough to be considered a weak perturbation to the total number of particles, such that the gamma-ray emission from the W44 interior is not strongly affected. To recover the same gamma-ray flux as in the case of unbroken shock, it is enough to increase the acceleration efficiency by 10-20\%.
In this scenario, the required masses for the two clouds are similar to the previous estimate, $M_{\rm NW}=5.4\times 10^3 \, M_{\odot}$ and $M_{\rm SE}=8.7\times 10^3 \, M_{\odot}$.

As we did for the Galactic CR model (see Sect. \ref{sec:model}), we calculated the $\chi^2$/N value using the best-fit model and including \fermilat up to 100 GeV + MAGIC data for NW source and \fermilat data up to 100 GeV for SE source. We found values of 33.6/28 and 23.1/22 for NW and SE, respectively, which confirms the good agreement between data and model.

Finally, we verified that the time needed for these particles to fill the two external regions has to be much shorter than the age of \w44. This time can be estimated as $t_{\rm fill} \simeq (2 R_{\rm SRC})^2/(2 D_{\rm ext}(p))$. For 1 GeV particles, it ranges between 4300\,yr and 2400\,yr for NW and SE sources, respectively. This corresponds to at most 20\% of the age of W44 and is consistent with the idea that the shock only starts to break in the very last phase of the SNR life.

\section{Summary}
\label{sec:summary}
We reported the results of \textit{Fermi}-LAT and MAGIC observations of the middle-aged SNR \w44 and its surrounding environment. 
Through the \textit{Fermi}-LAT high-energy analysis, we studied the morphology of the remnant in detail, finding that it resembles an ellipsoid. This confirms the results of previous studies. In particular, a radio template derived from the THOR survey provided the best description of the morphology of the SNR. 
Moreover, we confirmed the detection of two sources located at the NW and SE edges of the remnant shell. The extensions of the sources NW and SE are $r_{68}= 0\dotdeg16 \pm 0\dotdeg01$ and $r_{68}=0\dotdeg12 \pm 0\dotdeg01$, respectively. The smaller extension obtained for the NW source with respect to previous studies is probably due to the presence of a large diffuse source, which we modeled as a large central disk. 
We also derived the best spectral models, starting from 100 MeV, for \w44 SNR (described with an LP) and for the nearby sources (an LP for the large diffuse disk and a BPL for sources NW and SE).

The observations with the MAGIC telescopes focused on the northwestern region of the \w44. We carried out an exclusion region analysis considering as the position and extension of the NW source those derived from the \textit{Fermi}-LAT high-energy analysis. Since no significant detection was obtained, we derived 95\% CL ULs in the energy range above 100 GeV.

We note that no sources coincident with \w44, NW, or SE were identified in the first LHAASO catalog of gamma-ray sources \citep{Cao_2024} or in the third catalog of very high-energy gamma-ray sources of the High-Altitude Water Cherenkov gamma-ray Observatory (HAWC) \citep{albert_3hwc_2020}.

The combined \textit{Fermi}-LAT and MAGIC spectra of the NW source allowed us to reject a model in which this emission is entirely due to diffuse Galactic CRs.
We developed a model to describe the \w44, NW, and SE source emissions in the hypothesis of CRs that are freshly accelerated at the remnant shock and escape to the NW and SE clouds. This interpretation was based on the assumption that NW and SE sources are located at the same distance as \w44. However, to our knowledge, there are no other known sources along the same line of sight that might cause the observed NW and SE emissions.
The spectra of the \w44 SNR and the two close-by sources were described with a time-dependent model based on the assumption that CRs are accelerated at the SNR forward shock with a constant efficiency of $\sim 1\%$ up to a maximum energy that decreases in time as $E_{\max} \propto t^{-2}$. At the current age of \w44, the maximum energy is 44 GeV. At any given time, particles with energies $> E_{\max}(t)$ escape anisotropically from the forward shock, following the magnetic field lines along which the NW and SE sources are located. The emission from these sources was then interpreted as due to hadronic interactions of escaping protons with the ambient gas, whose total mass has to be $\approx 10^4\, \rm M_{\odot}$ for both sources. This model also requires that the diffusion coefficient around \w44 has to be suppressed by a factor of about 5 with respect to the average Galactic diffusion coefficient. We note that the NW and SE fluxes can be reasonably well described assuming the same CR acceleration efficiency as is required to fit the gamma-ray flux of \w44.
Finally, we note that the anisotropic CR escape was proposed in the past for another SNR, namely W28. However, for \w44, it is strongly supported by the recently measured direction of the large-scale magnetic field.

Gamma-ray data in the GeV range also indicate the presence of low-energy particles, down to $\sim 1$\,GeV, in NW and SE sources. At first glance, this  seems to contradict our escape model, in which only particles with $E \gtrsim 44$\,GeV can escape. However, we proposed a novel approach, the broken-shock scenario', to account for the escape of low-energy particles, in which a small fraction of the shock surface is destroyed (possibly by dense clumps), allowing the escape of particles at all energies. Compared to the total number of accelerated particles, the number of particles that escape from the broken -hock surface has to be $\approx 10\%$ and $20\%$ for SE and NWsources, respectively. This fraction is small enough for the acceleration process to not be significantly affected.
This model proved to agree reasonably well  with \textit{Fermi}-LAT spectral points and MAGIC ULs.

\begin{acknowledgements}
Contributions of the authors:
R.~Di~Tria: \fermilat analysis, paper drafting;
L.~Di~Venere: Project coordination, \fermilat analysis, paper drafting;
D.~Green: Project coordination, CO data analysis, paper drafting;
A.~Hahn: MAGIC analysis, paper drafting;
G.~Morlino: theoretical modeling and interpretation, paper drafting;
M.~Strzys: MAGIC analysis cross-check, paper drafting;
E.~Bissaldi: paper drafting;
S.~R.~Gozzini, A.~L\'opez-Oramas: PIs of MAGIC observation campaigns;
the rest of the authors have contributed in one or several of the following ways: design, construction, maintenance and operation of the instrument(s) used to acquire the data; preparation and/or evaluation of the observation proposals; data acquisition, processing, calibration and/or reduction; production of analysis tools and/or related Monte Carlo simulations; overall discussions about the contents of the draft, as well as related refinements in the descriptions.

\newline

The \fermilat Collaboration acknowledges generous
ongoing support from a number of agencies and institutes that have supported both the development and the operation of the LAT as well as scientific data analysis. These include the National Aeronautics and Space Administration and the Department of Energy in the United States, the Commissariat `a l’Energie Atomique and the Centre National de la Recherche Scientifique /Institut National de Physique Nucleaire et de Physique des Particules in France, the Agenzia Spaziale Italiana and the Istituto Nazionale di Fisica Nucleare in Italy, the Ministry of Education, Culture, Sports, Science and Technology (MEXT), High Energy Accelerator Research Organization (KEK) and Japan Aerospace Exploration Agency (JAXA) in Japan, and the K. A. Wallenberg Foundation, the Swedish Research Council and the Swedish National Space Board in Sweden. Additional support for science analysis during the operations phase is gratefully acknowledged from the Istituto Nazionale di Astrofisica in Italy and the Centre National d’Etudes Spatiales in France. This work
performed in part under DOE Contract DE-AC02-76SF00515.

%
\newline
We would like to thank the Instituto de Astrof\'{\i}sica de Canarias for the excellent working conditions at the Observatorio del Roque de los Muchachos in La Palma. The financial support of the German BMBF, MPG and HGF; the Italian INFN and INAF; the Swiss National Fund SNF; the grants PID2019-104114RB-C31, PID2019-104114RB-C32, PID2019-104114RB-C33, PID2019-105510GB-C31, PID2019-107847RB-C41, PID2019-107847RB-C42, PID2019-107847RB-C44, PID2019-107988GB-C22, PID2022-136828NB-C41, PID2022-137810NB-C22, PID2022-138172NB-C41, PID2022-138172NB-C42, PID2022-138172NB-C43, PID2022-139117NB-C41, PID2022-139117NB-C42, PID2022-139117NB-C43, PID2022-139117NB-C44 funded by the Spanish MCIN/AEI/ 10.13039/501100011033 and “ERDF A way of making Europe”; the Indian Department of Atomic Energy; the Japanese ICRR, the University of Tokyo, JSPS, and MEXT; the Bulgarian Ministry of Education and Science, National RI Roadmap Project DO1-400/18.12.2020 and the Academy of Finland grant nr. 320045 is gratefully acknowledged. This work was also been supported by Centros de Excelencia ``Severo Ochoa'' y Unidades ``Mar\'{\i}a de Maeztu'' program of the Spanish MCIN/AEI/ 10.13039/501100011033 (CEX2019-000920-S, CEX2019-000918-M, CEX2021-001131-S) and by the CERCA institution and grants 2021SGR00426 and 2021SGR00773 of the Generalitat de Catalunya; by the Croatian Science Foundation (HrZZ) Project IP-2022-10-4595 and the University of Rijeka Project uniri-prirod-18-48; by the Deutsche Forschungsgemeinschaft (SFB1491) and by the Lamarr-Institute for Machine Learning and Artificial Intelligence; by the Polish Ministry Of Education and Science grant No. 2021/WK/08; and by the Brazilian MCTIC, CNPq and FAPERJ.\\
We acknowledge the contributions of J.~Krause, V.~Stamatescu, and P.~Colin during the original observation campaigns.
\end{acknowledgements}
\bibliography{W44.bib}

\newpage

\begin{appendix}
\section{Galactic diffuse background contamination}
\label{sec:appendix}
\begin{figure}[h]
	\centering
	{\includegraphics[width=0.4\textwidth]{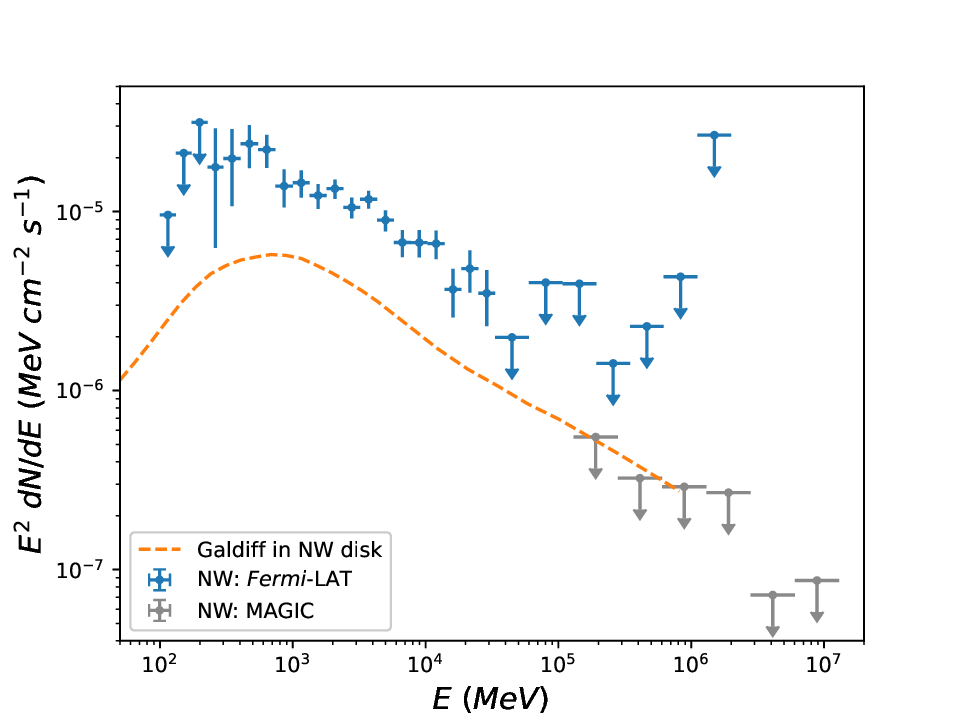}}
	{\includegraphics[width=0.4\textwidth]{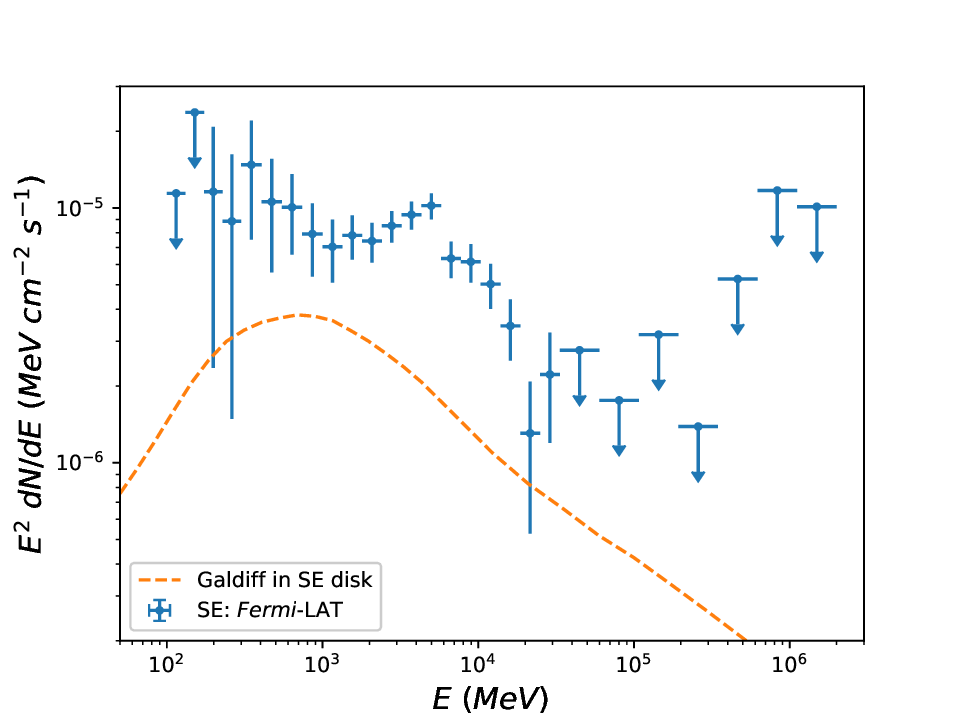}} 
	{\includegraphics[width=0.4\textwidth]{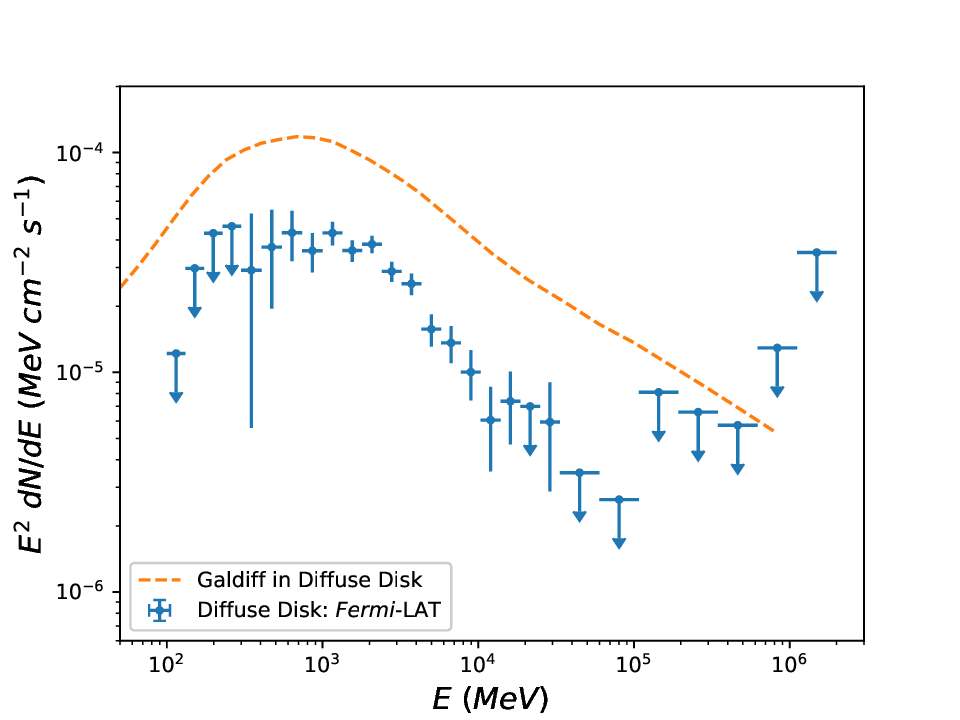}}
	\caption{Galactic diffuse background $E^{2}dN/dE$ (dashed orange line) evaluated in correspondence of the sources NW (top panel), SE (middle panel) and Diffuse Disk (bottom panel).}
	\label{fig:galdiff_seds}
\end{figure}

Here we report the $E^{2}dN/dE$ of the Galactic diffuse emission evaluated  from the model adopted for the \textit{Fermi}-LAT analysis \citep{Acero_2016} as described in Sect. \ref{sec:galdiff_analysis}. We considered three disks centered on the coordinates and having the same extensions of the NW (Fig.\ref{fig:galdiff_seds} top), SE (Fig.\ref{fig:galdiff_seds} middle) and diffuse disk (Fig.\ref{fig:galdiff_seds} bottom) sources. The SEDs of the three sources of interest are also shown in the same figures. Also in this case the errors are given by the quadrature sum of the statistical and systematic errors, the latter evaluated taking into account the uncertainties on the effective area.

\end{appendix}

\end{document}